\documentclass[iop]{revtex4-1}
\pdfoutput=1 
\pdfoutput=1
\expandafter\let\csname equation*\endcsname\relax
\expandafter\let\csname endequation*\endcsname\relax
\usepackage{mathtools} 
\usepackage[utf8x]{inputenc}
\usepackage{epsfig}
\usepackage{color}
\usepackage[toc,page]{appendix}
\usepackage{float}
\usepackage{graphicx}
\usepackage{anysize}
\usepackage{xfrac}
\usepackage{soul}
\soulregister\cite7
\soulregister\ref7
\soulregister\pageref7

\usepackage{soul}

\DeclareRobustCommand{\vect}[1]{
  \ifcat#1\relax
    \boldsymbol{#1}
  \else
    \mathbf{#1}
  \fi}

\marginsize{3cm}{3cm}{2.5cm}{4.5cm}
\begin{document}
\title[Effective Equations in complex systems: from Langevin to machine learning]{Effective Equations in complex systems: from Langevin to machine learning}
\author{A. Vulpiani}
\address{Department of Physics, Universit\`a ``Sapienza'', Roma  
Piazzale A. Moro 5, I-00185 (Italy)}
\address{Centro Interdisciplinare ``B. Segre'', 
Accademia dei Lincei, Via della Lungara 10, I-00165 Roma (Italy)}
\author{M. Baldovin}
\address{Department of Physics, Universit\`a ``Sapienza'', Roma  
Piazzale A. Moro 5, I-00185 (Italy).} 
\vspace{10pt}

\begin{abstract}
 The problem of  effective equations is reviewed and discussed.
Starting from the classical Langevin equation, we show how it can be generalized to 
Hamiltonian systems with non-standard kinetic terms. A numerical method for inferring
effective equations from data is discussed; this protocol allows to check the validity of our results.
In addition we show that, with a suitable treatment of time series, such protocol
can be used to infer effective models from experimental data. We briefly
discuss the practical and conceptual difficulties of a pure data-driven approach
in the building of models.
\end{abstract}

\maketitle

\section{ Introduction}

It is a matter of fact that  many interesting dynamic problems in science and 
engineering are characterized by the presence of 
 a variety of degrees of freedom with very different time scales. 
As  important examples we can mention proteins~\cite{karplus76} and climate dynamics~\cite{donges09}:
we remind that  the time scale of vibration of covalent bonds is 
$O(10^{-12} s) $, while the folding time for proteins may be of the order 
of seconds;
in a similar way  the characteristic times of the  processes
involved in climate  vary from  seconds
 for the $3D$ turbulence,  to days for the atmosphere, to   $O(10^4 yr)$
 for the deep ocean currents and ice shield dynamics.
\\
Due to the multi-scale character of such kind of systems,
is not possible to perform  a direct  simulation
of  all the relevant  involved scales, 
even with the support of modern supercomputers  and advanced
numerical algorithms. 
These practical difficulties force us to reduce our ambitions;
a (non-trivial) possibility in this sense is to describe  the  "slow dynamics" in terms 
of effective equations.
Using such an approach  one  has  both practical and conceptual advantages:
for instance, it is possible to decrease the computational effort, e.g. by reducing the number of equations and adopting a "large" $\Delta t$;
in addition, the effective equations are able to catch some general features and to reveal dominant ingredients which can remain hidden in the 
detailed description~\cite{tnt_entropy}.
\\       
Disappointingly enough, only in few cases it is possible to derive effective equations with a systematic approach: important examples are dilute gases \cite{dorfman72},
 harmonic chains~\cite{rubin60, zwanzig73} and the Markovian limit of Hamiltonian dynamics \cite{spohn80}.
\\
On the other hand, in the history of science there is a series
of clever practical approaches for the study of multi-scale problems
that do not rely on rigorous derivations, e.g. the averaging method in  celestial mechanics~\cite{mitropolsky67},
the Langevin equation for colloids~\cite{langevin},
the homogenization for partial differential equations~\cite{pavliotis08},
the Born-Oppenheimer "approximation"~\cite{hutter09} and
the Carr-Parrinello method~\cite{car85}.
\\
In order to give an idea of the general methodology
let us briefly remind a  well known example of effective model,
the advection-diffusion equation for a passive scalar $\rho({\bf x},t)$ 
(e.g. the concentration of a pollutant) 
in an incompressible flow $(\nabla \cdot {\bf u}=0)$:
\begin{equation}
 \label{eq:adv-diff}
{\partial \rho \over \partial t}+({\bf u} \cdot \nabla) \rho=
D_0 \Delta \rho \,\, .
\end{equation}
Maxwell had the idea, now  supported 
 by mathematics (under rather general conditions), to consider the solution of Eq.~\eqref{eq:adv-diff}
at large scale and asymptotically in time; in these limits one obtains the so called standard diffusion,
i.e. a  Fick's law holds of the form
\begin{equation}
 {\partial \Theta \over \partial t}=
\sum_{i,j} {\cal D}_{ij} {\partial^2 \Theta \over \partial x_i \partial x_j}
\end{equation}
where $\Theta$ is the spatial coarse graining of $\rho$, and ${\cal D}_{ij}$ is the
effective (eddy) diffusion tensor, depending (often in a non trivial way)
on $D_0$ and the field ${\bf u}$
(just for simplicity we considered the case $\langle{\bf u}\rangle=0$).
If some (rather general) conditions on  the field ${\bf u}$ are satisfied then
one can use a precise protocol to compute the tensor ${\cal D}_{ij}$~\cite{biferale95}.
\\
In this paper we review some important aspects of the problem of finding effective
equations for complex systems. In Section~\ref{sec:generalizing} we review the classical
Langevin equation and show how it can be extended to cases in which the system
obeys a Hamiltonian with a generalized form of the kinetic energy; Section~\ref{sec:testing}
is devoted to the discussion of a data-driven method that allows to test such generalization;
Section~\ref{sec:granular} shows how this method can be applied to experimental cases, and how it can be used to
infer coarse-grained models whose behavior nicely agree with that of the real system; then in Section~\ref{sec:bigdata}
we briefly comment on the lessons that we can learn from the problem of effective equations in physics,
and how such warnings can reveal useful also in the context of big data and machine learning. In Section~\ref{sec:conclusion}
we briefly sketch our conclusions.

\section{Generalizing the Langevin Equation}
\label{sec:generalizing}
In his celebrated paper about Brownian motion, Paul Langevin addressed the problem of properly describing the irregular behavior of pollen particles suspended in water~\cite{langevin}. Following Einstein, he assumed that 
both the colloidal particle and the molecules of the fluid could be modeled as material points with masses $M$ and $m\ll M$ respectively. The motion of the heavy particle is due to the collisions with the molecules of the liquid, which are assumed to be uncorrelated. To account for the discontinuous action of the hitting molecules, Langevin relied upon the introduction of a stochastic term in the evolution equation of the colloid, namely a white Gaussian noise $\boldsymbol{\xi}(t)=(\xi_1(t),\xi_2(t),\xi_3(t))$ such that 
\begin{equation}
\begin{aligned}
 \langle \xi_i (t) \rangle&=0\\
 \langle \xi_i (t) \xi_j (t') \rangle&=\delta(t-t')\delta_{ij}\,.
\end{aligned}
\end{equation}
He supposed that the impulsive force acting on the colloid was proportional to this discontinuous function $\vect{\xi}$.
On the other hand, he argued that the interaction with the fluid results into an average damping force acting on the colloidal particle, and proportional to its velocity (Stokes law). The combination of the above effects leads to the celebrated Langevin Equation~(LE)
\begin{equation}
\label{eq:LE}
 \dot{\vect{P}}=-\gamma \vect{P}+\sqrt{2 \gamma k_B T} \vect{\xi}
\end{equation}
which characterizes the evolution of the momentum $\vect{P}=M\dot{\vect{Q}}$ of the heavy colloid ($\vect{Q}$ being its position in the three-dimensional space). Here $\gamma$ is the friction term due to the interaction of the colloidal particle with the fluid, while $T$ is the temperature and $k_B$ is the Boltzmann constant.
The noise amplitude is determined by the Einstein relation, which relies on the fact that the interested particle is at thermal equilibrium with the fluid, so that equipartition theorem holds and
\begin{equation}
\label{eq:equipartition}
 \langle P^2/M \rangle =3 k_B T\,.
\end{equation}
\\
Eq.~\eqref{eq:LE} clearly shows that the Brownian motion is the result of the competing actions of a damping force and a thermal noise. It seems reasonable that such mechanism should hold, under appropriate modifications, also for Hamiltonian systems with non-standard, generalized forms of the kinetic energy $K(P)$ (e.g., non quadratic functions of the momenta). In these cases there would be no reason for the damping force to be proportional to the momentum, and the equipartition theorem could assume a formulation very different from Eq.~\eqref{eq:equipartition}.
\\
This problem has been addressed in~\cite{baldovin18} and, as it will be discussed in the following, it assumes particular conceptual relevance when Hamiltonian systems living in bounded phase-spaces are taken into account, so that the absolute temperature of the system can assume negative values.
\\
Let us consider the general case of a Hamiltonian system in the form
\begin{equation}
\label{eq:ham_gen}
  H(P,\{p_n\},Q,\{q_n\})=K(P)+\sum_n \tilde{K}(p_n)+ U(Q) + V(Q,\{q_n\})
 \end{equation}
 where $(P,Q)$ is a ``slow'' degree of freedom. For example, in a system with the usual quadratic kinetic energy it could represent a particle with a mass much higher than the others ($K(P)=P^2/2M$, $\tilde{K}(p_n)=p_n^2/2m$, $M\gg m$). $U(Q)$ is the external potential which the slow particle is subjected to, while $V(Q,\{q_n\})$ takes into account the interactions occurring among different degrees of freedom. For the sake of simplicity we consider here only Hamiltonian systems in one dimension, but all the results can be straightforwardly generalized to the multi-dimensional case.
 \\
 In what follows we will limit our analysis to Hamiltonians of the form~\eqref{eq:ham_gen}, in which the kinetic energy is the sum of single-particle contributions only depending on the momentum.
 \\
 The Hamilton equations describing the motion of the slow degree of freedom read:
 \begin{equation}
 \label{eq:motion}
  \begin{cases}
   \dot{Q}=\partial_PK(P)\\
   \dot{P}=-\partial_QU(Q)-\partial_QV(Q,\{q_n\})
  \end{cases}
 \end{equation}
 At this stage we introduce a first, strong hypothesis, in the spirit of the one done by Langevin: we suppose that the time-scale separation between the dynamics of the slow particle and the fast ones allows us to approximate the former through an effective stochastic equation. In other words, we assume that Eqs.~\eqref{eq:motion} can be rewritten as 
 \begin{equation}
 \label{eq:kleinkramers}
  \begin{cases}
   \dot{Q}=\partial_PK(P)\\
   \dot{P}=-\partial_QU(Q)+\Gamma(P)+\sqrt{2D}\xi(t)\,.
  \end{cases}
 \end{equation}
  Here the term $\Gamma(P)$ can be seen as a generalization of the Stokes force, while $D$ is a constant which determines the amplitude of the noise. In this way we are ignoring the details of the interactions between the slow and the fast degrees of freedom. The possibility to perform such averaging procedure on rigorous mathematical grounds is a non-trivial, largely studied problem in the field of dynamical systems~\cite{givon04, kifer04}: the above approximation should be therefore viewed as an ansatz, whose validity needs to be checked \textit{a posteriori}. 
 \\
 We aim at finding some kind of generalized Einstein relation to relate the constant $D$ to $\Gamma(P)$. Let us introduce the steady probability density $f(Q,P)$ of the considered degree of freedom (to be determined) and the corresponding steady probability currents:
\begin{equation}
\begin{cases}
 J_Q(Q,P)=f(Q,P)\partial_PK(P)\\
 J_P(Q,P)=-f(Q,P)\partial_QU(Q,P)+\Gamma(P)f(Q,P)-D\partial_P f(Q,P)\,. 
\end{cases}
\end{equation}
In terms of the above quantities, the Fokker-Planck equation corresponding to Eq.~\eqref{eq:kleinkramers} reads:
\begin{equation}
\label{eq:fokkerplanck}
 \partial_QJ_Q(Q,P)+\partial_PJ_P(Q,P)=0\,.
\end{equation}
\\
We assume now that the system is in thermal equilibrium (which is the same hypothesis done by Einstein and Langevin when 
exploiting the equipartition theorem~\eqref{eq:equipartition}). We require therefore that detailed balance is satisfied, i.e. that the irreversible part of $J_P$ vanishes, so that
\begin{equation}
\label{eq:gamma}
\Gamma(P)f(Q,P) -D\partial_Pf(Q,P)=0\,. 
\end{equation}
Exploiting the factorization of the equilibrium distribution $f(Q,P) = f_Q(Q) f_P(P)$ and the fact that
\begin{equation}
 f_P(P)\propto e^{-\beta K(P)}
\end{equation}
where $\beta=(k_B T)^{-1}$, one easily finds from Eq.~\eqref{eq:gamma}:
\begin{equation}
\label{eq:einstein_gen}
   \Gamma(P)=-D \beta \partial_P K(P)\,.
\end{equation}
This last equation can be seen as a generalization of the Einstein relation to cases with non-quadratic kinetic energy. It tells that the Stokes law is always proportional to the velocity $\dot{Q}=\partial_P K(P)$, no matter what the form of the kinetic energy is, and that their ratio is fixed by $-D\beta$.
Let us stress that in the usual, Newtonian case $K(P)=P^2/2M$, the Einstein relation is exactly recovered, as it should.
\\
The above argument gives a relation between $D$ and $\Gamma$, but it is not sufficient to determine $\Gamma$ (or $D$) from the knowledge of the Hamiltonian. When the nature of the bath is specified one may try perturbative methods in the limit of large scale separation to derive all the parameters of the effective equation. An analytically tractable case has been discussed in Ref.~\cite{baldovin19}, where the thermal bath is constituted by a large number of Ising spins, which are kept at a fixed temperature by a Glauber dynamics, and the ``slow'' degree of freedom is an oscillator with generalized kinetic energy.
All the spins feel a magnetic field that depends on the position of the oscillator. In the limit in which the typical frequency of the oscillator $\omega_{osc}$ is much slower than the rate of the Glauber dynamics $r_{spin}$, a Chapman-Engsok expansion of the Fokker-Planck equation of the particle can be performed, for which the small parameter is given by
\begin{equation}
\epsilon=\frac{\omega_{osc}}{r_{spin}}\,. 
\end{equation}
\\
The obtained  Langevin equation for the slow dynamics is of the form
\begin{equation}
\label{eq:prados_final}
\begin{aligned}
 \dot{Q}&= K'(P)\\
 \dot{P}&=-U'_R(Q) - \Gamma(P,Q) K'(P) +\sqrt{2 D(Q)} \eta
\end{aligned}
\end{equation}
where  $U_R(Q), \Gamma(P,Q)$ and  $D(Q)$ can be explicitly computed. 
Remarkably, this result basically coincides with the one obtained with  the previous 
phenomenological argument: Eq.~\eqref{eq:einstein_gen} still holds, with the only
difference that $D$ is now a function of $Q$.
\\
It can be verified that both equilibrium features (e.g. stationary probability density function), as well as non-equilibrium  ones
(e.g. correlations and relaxations) obtained with  the effective Langevin equation~\eqref{eq:prados_final}
are in perfect agreement  with the actual numerical  results from the complete system~\cite{baldovin19}.

\section{Testing the generalized LE}
\label{sec:testing}
\begin{figure}[!h]
  \includegraphics[keepaspectratio=true, width=0.48\linewidth]{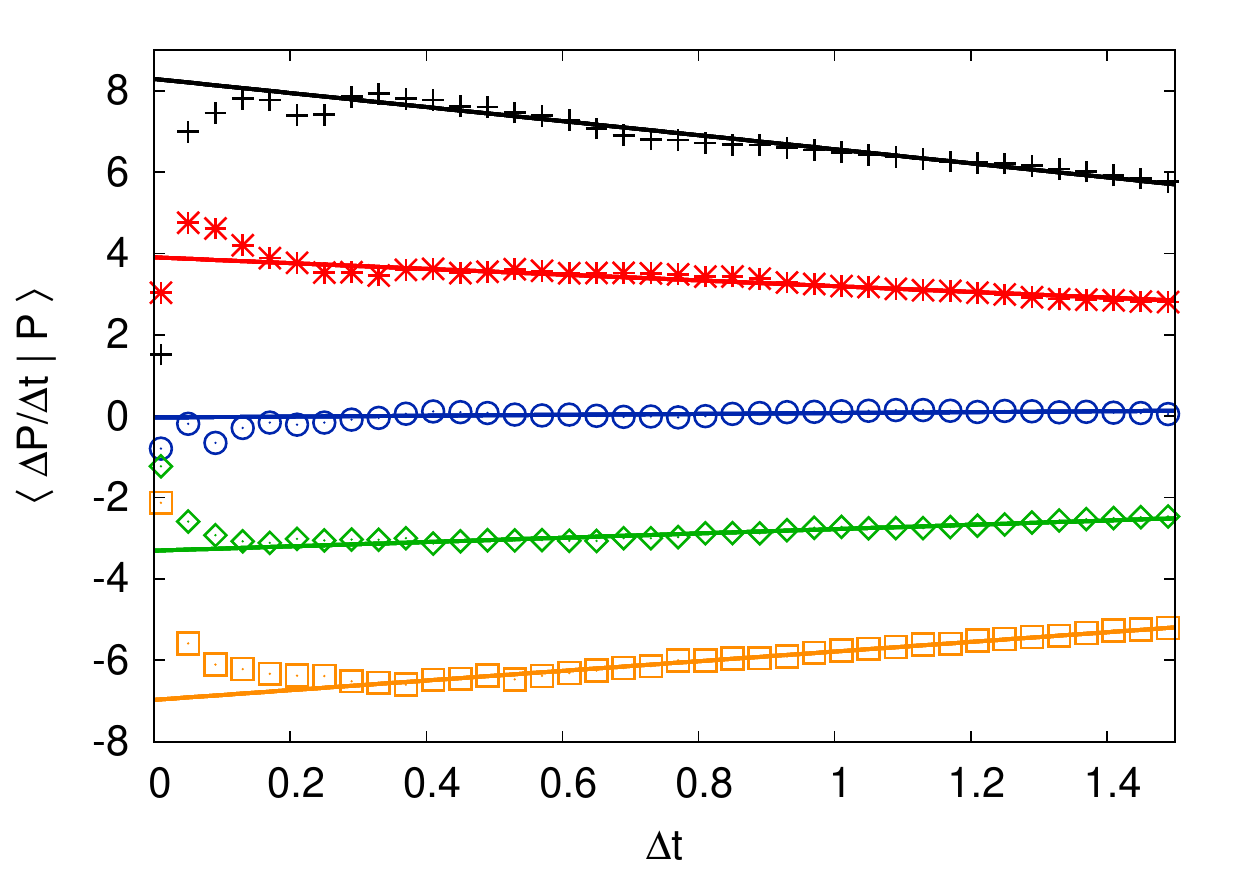}
  \includegraphics[keepaspectratio=true, width=0.48\linewidth]{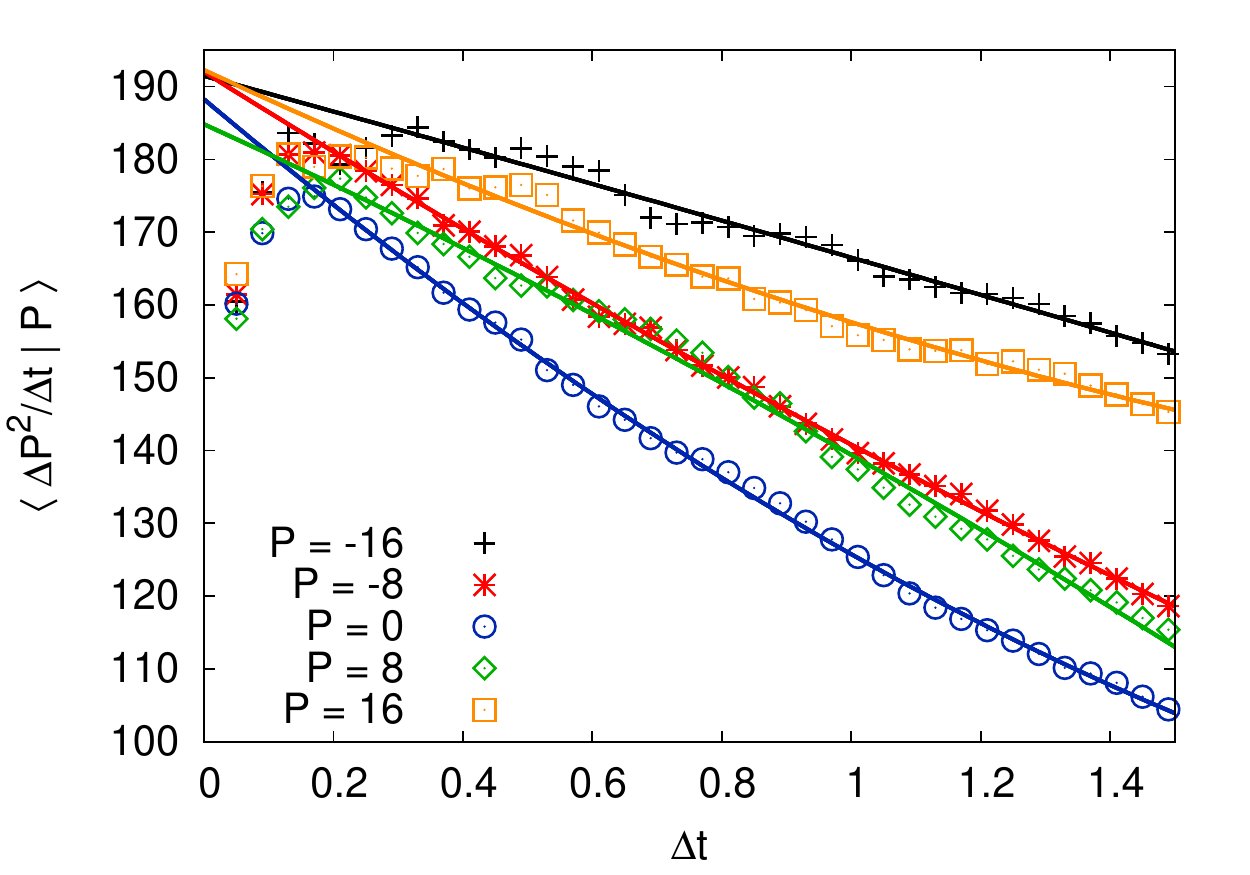}
 \caption{Evaluation of the conditioned moments on the r.h.s. of Eq.~\eqref{eq:d&d}. We numerically compute such quantities as functions of $\Delta t$ (points), then we fit the curves with low-order polynomials (solid lines) and we consider the limits for $\Delta t \to 0$. Left: r.h.s. of Eq.~\eqref{eq:d&d}a, linear fit. Right: r.h.s. of Eq.~\eqref{eq:d&d}b, parabolic fit. Different colors and shapes of the points correspond to different values of the initial value of $P$. All fits are computed between $\Delta t=0.25$ and $\Delta t=1.5$. Parameters of the simulation  $M=200$, $m=1$, $k=2500$, $2N=2000$, $\beta\simeq1.0$. }\label{fig:d&d}
\end{figure}

\begin{figure}[!h]
\centering
   \includegraphics[keepaspectratio=true, width=0.7\linewidth]{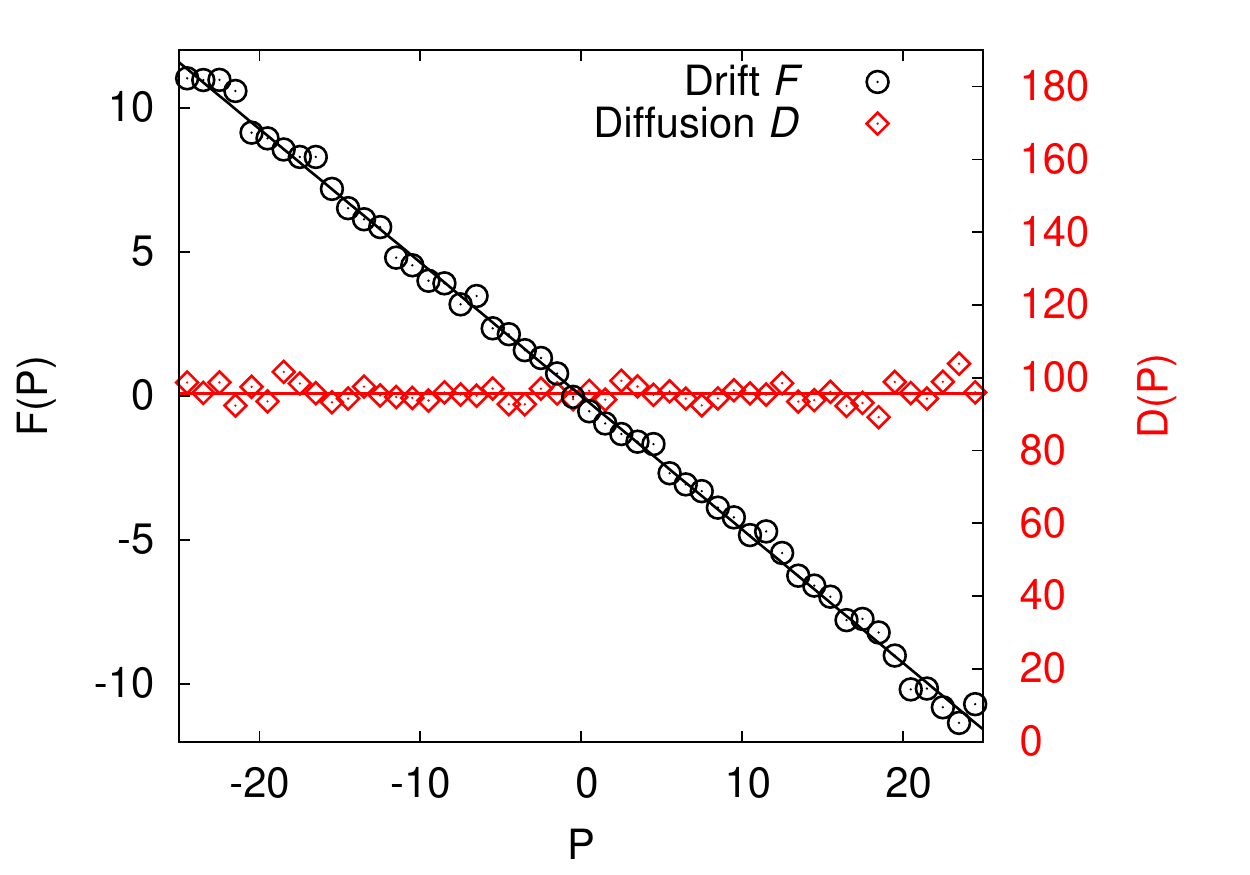}
   \caption{Drift term $F(P)$ and diffusivity $D(P)$ of the process $P(t)$ as determined by the data-driven procedure discussed in the text. Red circles (black diamonds) represent the values obtained for the drift (diffusivity) from the limits~\eqref{eq:d&d}; solid lines are linear fits.}\label{fig:fit_osc}
\end{figure}

\begin{figure}[!h]
\centering
 \includegraphics[keepaspectratio=true, width=0.48\linewidth]{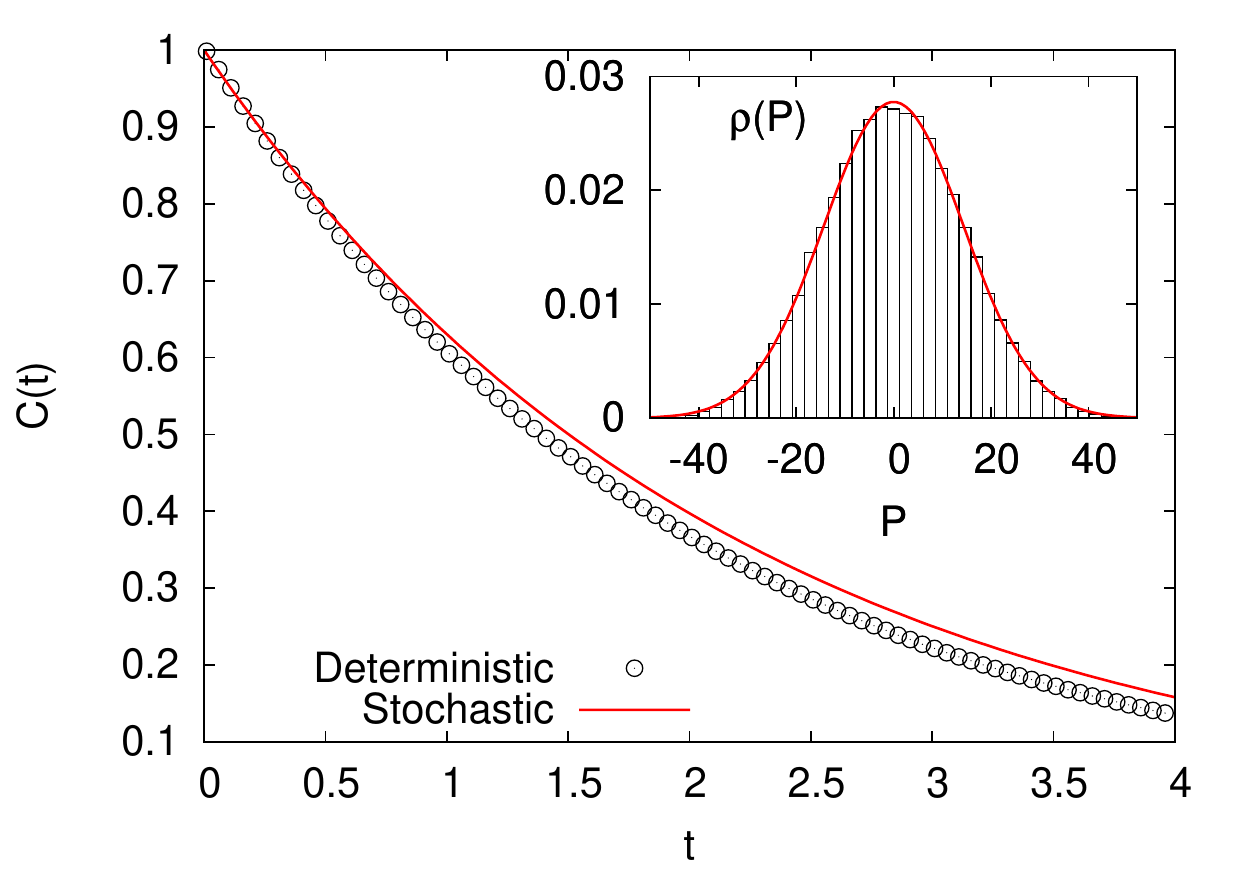}
  \includegraphics[keepaspectratio=true, width=0.48\linewidth]{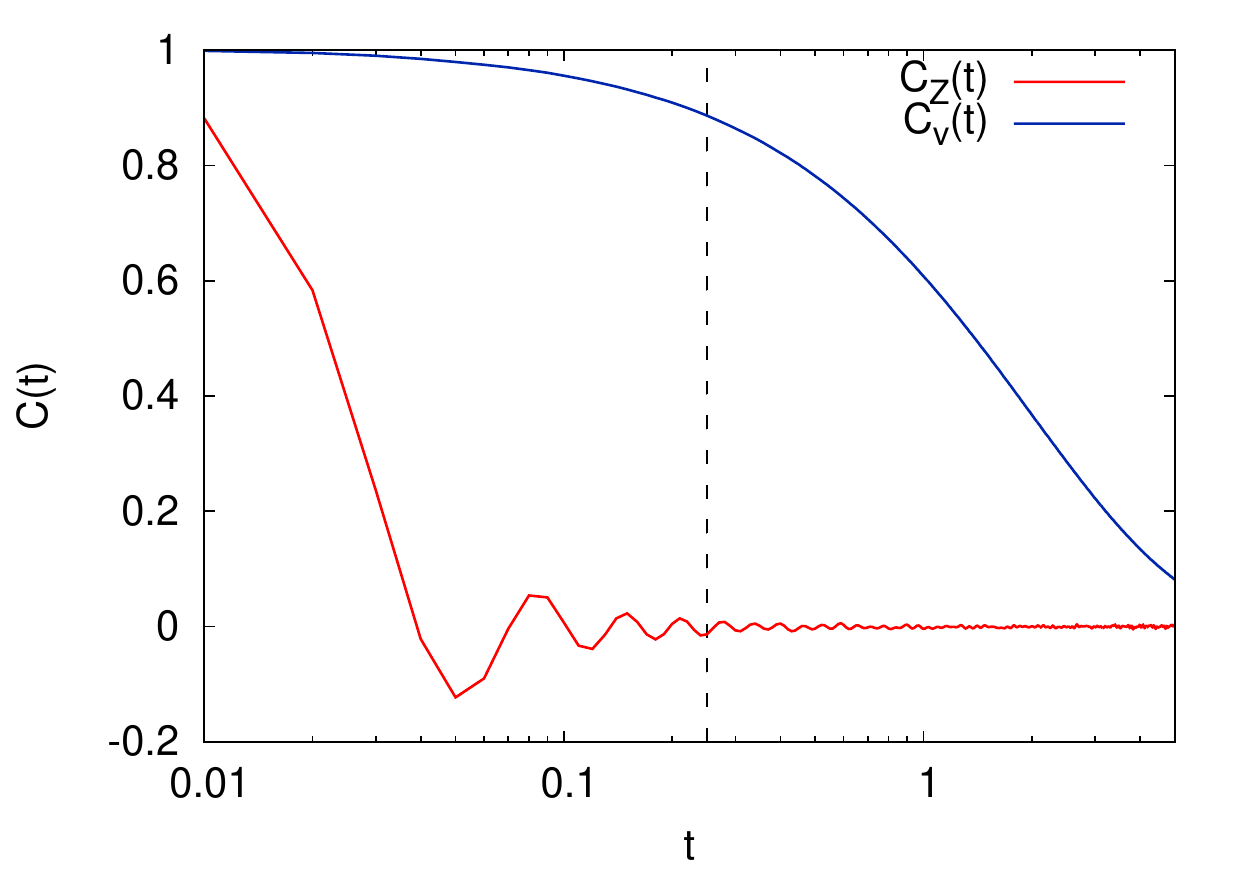}
 \caption{Left: autocorrelation function of the velocity in the original dynamics (circles) and in the reconstructed stochastic process (solid line). In the inset, the steady probability density functions for the deterministic evolution (boxes) and for the stochastic one (solid red line) are compared. Right: autocorrelation function of the velocity (blue) compared to the one of the ``thermal noise'' defined by Eq.~\eqref{eq:zeta} (red).}\label{fig:autoc_osc}
\end{figure}

In order to check the validity of the generalized form for the LE,
\begin{equation}
\label{eq:gl}
 \dot{P}= -\partial_QU(Q) -D \beta \partial_P K(P) + \sqrt{2D} \xi\,,
\end{equation}
we can perform computer simulations of a large, compound system in which both ``heavy'' and ``light'' particles are present, and compare the effective behavior of one slow particle with the stochastic description given by Eq.~\eqref{eq:gl}. Our strategy articulates into four steps:
\begin{enumerate}
 \item Design a suitable Hamiltonian system in which time-scale separation may be expected;
 \item Simulate a (deterministic) evolution of such system;
 \item Extrapolate \textit{a posteriori} the coefficients of the effective LE which approximates the dynamics to the best extent;
 \item Compare them to Eq.~\eqref{eq:gl}.
\end{enumerate}
First, we will carry out our program with a Hamiltonian system which is well-known to reproduce the Brownian motion in the thermodynamic limit, i.e. a harmonic chain with an heavy ``intruder'':
 \begin{equation}
 \label{eq:ham_osc}
 H=\frac{P^2}{2M} + \sum_{i=\pm 1, ..., \pm N} \frac{p_i^2}{2m} +  \frac{k}{2}\sum_{i=-N}^{N+1} (q_i-q_{i-1})^2, \quad \quad Q \equiv q_0\,.
  \end{equation}
  Here $(p_i,q_i),\,i=-N,...,-1,1,...,N$ are the canonical coordinates of the ``light'' particles, with equal masses $m$, while $(P,Q)$ are those of the heavy intruder of mass $M\gg m$; $k$ is the elastic constant. We consider fixed boundary conditions $q_{-N-1}=q_{N+1}=0$.
The above model and similar harmonic chains have been analytically studied since the 1960's and represent one of the few examples in which stochastic differential equations can be exactly derived starting from first principles~\cite{rubin60, turner60, ford-kac-mazur65, zwanzig73}.
\\
Hamiltonian~\eqref{eq:ham_osc} is integrable, so that the energy assigned to each normal mode at the beginning of the dynamical evolution is conserved; as a consequence, if the system is initialized in such a way that energy is shared among only few degrees of freedom, thermodynamic equilibrium will never be reached and the Langevin description~\eqref{eq:LE} will necessarily fail. If, conversely, the system starts at equilibrium, it can be rigorously shown that the dynamics of $P$ is approximated by a Markovian stochastic process, whose autocorrelation function reads
\begin{equation}
C(t) \simeq \exp\left(-\frac{2\sqrt{k m}}{M} t \right) + O(m/M)\,.
\end{equation}
\\
We numerically simulate Hamiltonian~\eqref{eq:ham_osc} with a standard velocity Verlet update, choosing the time-step in such a way that the relative fluctuations on the total energy are of order $O(10^{-5})$. We start from equilibrium initial conditions.
\\
Given a generic Langevin Equation
\begin{equation}
 \dot{P}=F(P)+\sqrt{2 D(P)}\xi
\end{equation}
the drift term $F(P)$ and the diffusivity $D(P)$ can be computed from the temporal evolution of $P$ using the definitions~\cite{gardiner85}
 \begin{subequations}
 \label{eq:d&d}
   \begin{align}
  F(P)&=\lim_{\Delta t \to 0}\, \frac{\langle \Delta P | P(t_0)=P \rangle}{\Delta t}\\
  D(P)&=\lim_{\Delta t \to 0}\, \frac{ \langle \Delta P^2 | P(t_0)=P \rangle}{2 \Delta t}\,.
  \end{align}
 \end{subequations}
 In other words, we can evaluate the Langevin coefficients for a given value $P$ of the variable by looking at the average behavior of the trajectory after it passes through $P$. 
This approach has been used in several contexts ranging from physics to biology and finance~\cite{peinke11, peinke18}.
\\
Of course, the above limits of conditioned moments need to be evaluated with care. One has to be sure that the sampling rate is much higher than the typical frequencies of the dynamics, so that the quantities on the r.h.s. of Eq.~\eqref{eq:d&d} can be evaluated for time intervals $\Delta t$ smaller than any characteristic time of the evolution. In this case, in particular, one needs $\Delta t \ll M/\sqrt{k m}$.
\\
On the other hand, the evolution cannot be Markovian on all time-scales, because the original dynamics is deterministic, and it depends on the interactions with other degrees of freedom. This can be also understood by considering the velocity autocorrelation functions $C_D(t)$ and $C_L(t)$ in a deterministic and in a Langevin process respectively; for small times $t$, they can be expanded as:
\begin{subequations}
\begin{align}
 C_D(t)=1- {t^2 \over \tau_D^2}+O(t^3)\\
 C_L(t)=1- {t \over \tau_L}+O(t^2)\,.
\end{align}
\end{subequations}
Therefore it exists a minimal time-scale $\tau_M$ (sometimes called the Markov-Einstein time~\cite{peinke11}), such that the process can be considered Markovian only on time-scales much larger than $\tau_M$. Such threshold should be at least $O\Bigl({\tau_D^2 \over \tau_L}\Bigr)$, in order for the differences between $C_D$ and $C_L$ to be negligible.
\\
At a practical level, a good strategy consists in evaluating the quantities~\eqref{eq:d&d}a and ~\eqref{eq:d&d}b (for a fixed starting value $P$) as functions of the time interval $\Delta t$, then looking at their behavior for $\Delta t \le \tau_M$ (but still small with respect to the typical times of the evolution) and extrapolating the limit $\Delta t \to 0$ (Fig.~\ref{fig:d&d}).
In order to numerically evaluate the conditioned moments relative to an initial value, say, $P_0$, we have to wait until the trajectory passes through a (small) interval $(P_0-\Delta, P_0+\Delta)$, and then to look at its displacement $\Delta P$ after a time $\Delta t$. This is repeated many times, so that the averages in Eq.~\eqref{eq:d&d} are evaluated as temporal averages.
\\
The above procedure is performed for several values of the starting value of $P$, so that at the end we can appreciate the dependence of the drift and the diffusivity on $P$. The results are shown in Fig.~\ref{fig:fit_osc}.
As expected, the drift linearly depends on the momentum and the diffusivity is constant. Relation~\eqref{eq:einstein_gen} is also verified.
In order to check that the reconstructed LE actually reproduces the behavior of the slow particle, we can do an additional check: we can compute the steady probability density and the autocorrelation function in this new coarse-grained dynamics and compare them to the original, deterministic evolution. In this simple case, since the dynamics is linear, we can determine such observables analytically once we know $F(P)$ and $D$; in more complicated cases one can rely on numerical simulations of the stochastic process. Fig.~\ref{fig:autoc_osc}(a) shows both quantities in the original and in the reconstructed dynamics: the agreement is quite good.
Finally, we have to check that {time-scale separation} hypothesis is valid, i.e. that the ``thermal noise''
\begin{equation}
\label{eq:zeta}
  \zeta(t)=\dot{P}(t)-F(P(t))
\end{equation}
 decorrelates much faster than  $P$. The autocorrelation functions of the two quantities are shown in Fig.~\ref{fig:autoc_osc}(b): the time-scale separation is evident.

\subsection{A case with also negative temperature}
According to Statistical Mechanics, the inverse temperature $\beta$ 
is defined  in terms of the microcanonical entropy $S(E)$
of the system described by $H(\mathbf{X})$:
\begin{equation}
\label{eq:beta}
 \beta=\frac{1}{k_B}\frac{\partial S}{\partial E},\quad \quad
 S(E)=k_B \ln\int d{\mathbf{X}}\, \delta(E-H(\mathbf{X})) \, .
\end{equation}
\\
From the above expression one realizes that
 $\beta$ becomes negative if $S(E)$ is a decreasing function. It is well known
 that this never happens for systems with the usual, quadratic kinetic energy~\cite{ramsey56},
 so that negative temperatures can be only observed for peculiar systems,
 typically living in bounded phase-spaces.
\\
Systems with negative temperature, however, are not mere curiosities:
among the many interesting physical cases we can mention
$2D$ inviscid hydrodynamics, including the point-vortices model discussed
by Onsager during the first StatPhys conference (Florence 1949)~\cite{onsager49}, systems of nuclear spins~\cite{ramsey56},
the discrete nonlinear Schr\"odinger equation~\cite{iubini13} and systems of cold atoms trapped in optical lattices~\cite{braun13}.
\\
Since in our derivations we never assumed that the temperature of the system is positive, it can be interesting to see if formula~\eqref{eq:einstein_gen} still holds for Hamiltonian models which can assume negative temperature.
We will consider the following form for the kinetic energy:
\begin{equation}
 K(p)=mc^2\left[1-\cos \left(\frac{p}{mc}\right)\right]\,.
\end{equation}
Here $c$ is a constant with the physical dimensions of a velocity, while $m$ can be seen as a generalized ``mass'': it is straightforward to verify that, at fixed velocity, both kinetic energy and momentum are proportional to $m$, as one would expect from additivity.
Momentum $p$ lives in the interval $[-mc \pi, mc \pi)$, so that it can be considered as an angular variable. If also the ``positions'' are chosen to live in a bounded space, we can expect to observe negative temperature at high values of the energy.
Apart from the constant, the above form of the kinetic energy resembles the one that has been observed in a famous experiment on cold atoms~\cite{braun13}, and it has been used in the definition of mechanical models for systems with negative temperature~\cite{hilbert14, cerino15, baldovin17, miceli19}. One of the reasons is that for small energies such term reduces to the usual, quadratic form $p^2/2m$, so that in the low (positive) temperature regime we recover the usual statistical properties.
\\
Also in this case, we want to study the effective motion of a ``heavy'' particle subjected to the action of a thermal bath of ``light'' molecules. We need that the intruder interacts simultaneously with many, uncorrelated, degrees of freedom, in such a way that it can be considered at thermal equilibrium with such reservoir. In the following, we will model the bath as a chain of ``oscillators'' with equal masses $m$:
 \begin{equation}
  H_{chain}=\sum_{i=1}^{N}mc^2\left[1-\cos \left(\frac{p_i}{mc}\right)\right]+\epsilon \sum_{i=1}^{N+1}[1-\cos(\theta_i-\theta_{i-1})]
 \end{equation}
where $(p_i,\theta_i)$ is the $i-$th pair of canonical coordinates.
We coupled then such chain to a ``heavy'' degree of freedom $(\Theta,P)$ of mass $M\gg m$ via a bounded potential, so that the total Hamiltonian reads
  \begin{equation}
  H=H_{chain}+Mc^2[1-\cos(P/cM)] + k \sum_{i=1}^{N/n} [1-\cos(\Theta-\theta_{i\cdot n})]\,.
 \end{equation}
Let us stress that the heavy particle is only coupled to those oscillators whose label is a multiple of the integer parameter $n\gg 1$.
\\
We can expect that the different inertia of the intruder makes its dynamics much slower than that of the other degrees of freedom, so that a time-scale separation should be observed. Moreover, our choice of the interaction potential should keep the particle at the same temperature of the thermal bath. Remarkably enough, in this case such temperature can be negative, due to the bound on the total phase space volume. If the generalized Einstein relation~\eqref{eq:einstein_gen} holds for $\beta<0$, it means that $F(P)$, in such regime, must be positive when $P>0$ and negative when $P<0$, contrary to what happens at positive temperature. In other words, the (generalized) Stokes force tends to give energy to the particle, instead of subtracting it. At first sight this behavior may appear very unphysical: it can be understood by remembering that a thermal bath at negative temperature increases its entropy by decreasing the internal energy, i.e. by releasing heat.
\\
We can apply to this new model the data-driven analysis discussed in the previous section.
\begin{figure}
 \centering
  \includegraphics[keepaspectratio=true, width=0.49\linewidth]{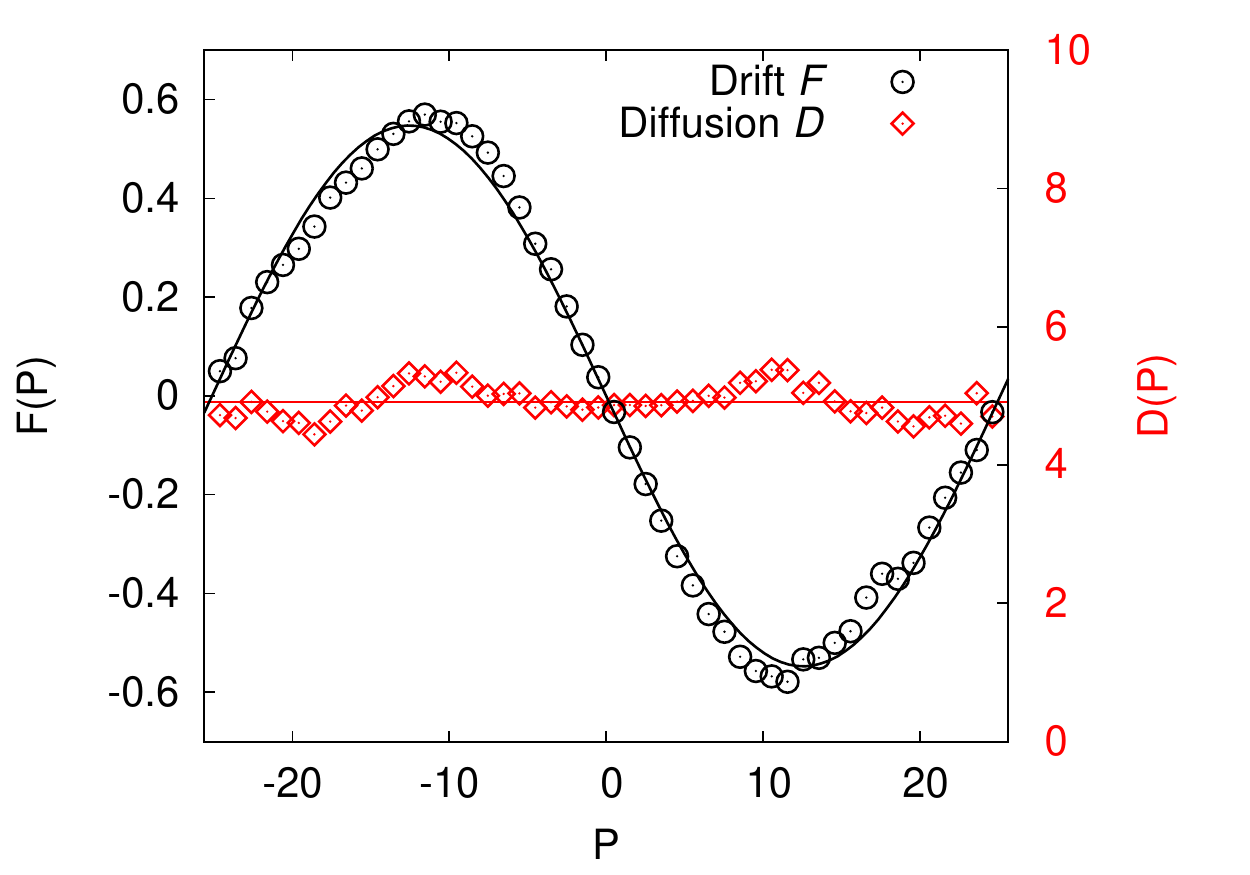}
  \includegraphics[keepaspectratio=true, width=0.49\linewidth]{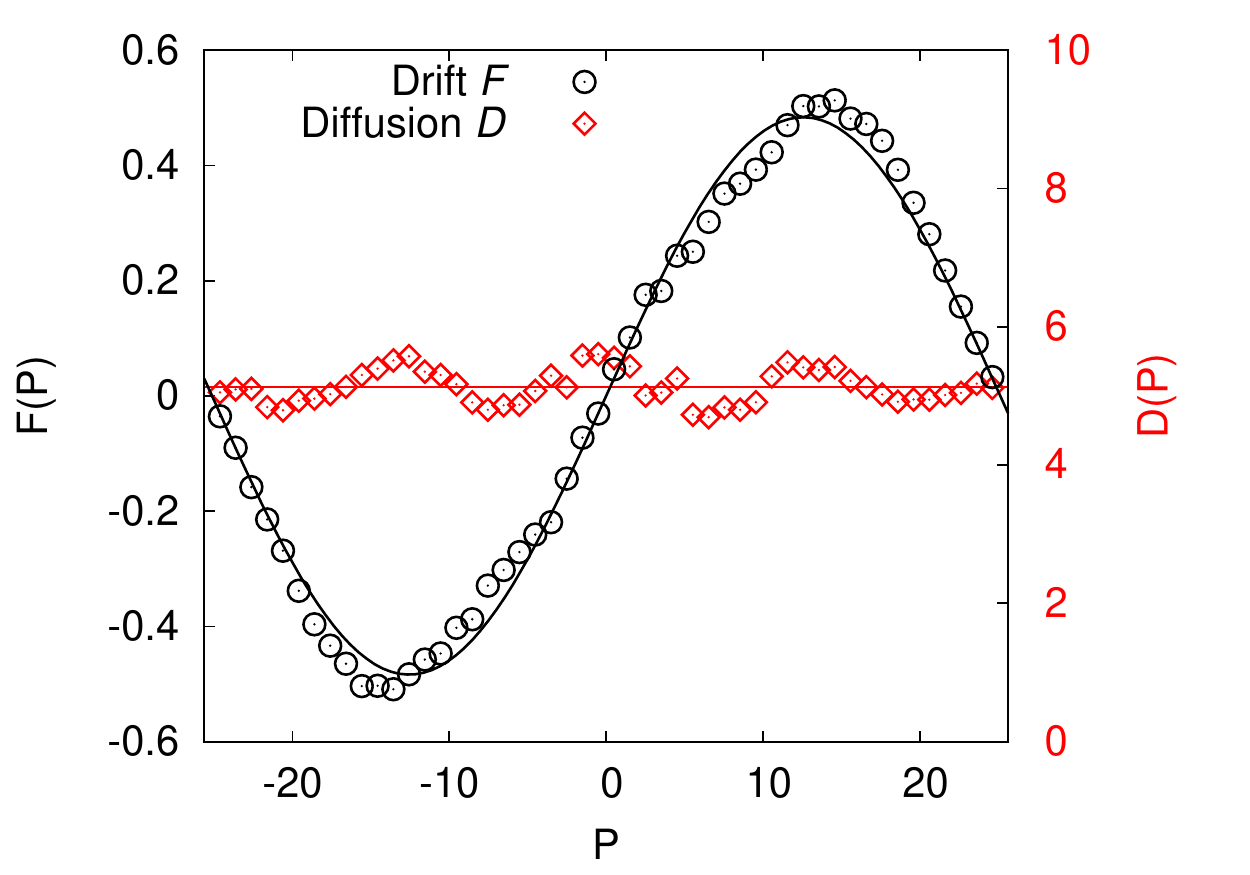}
  \caption{Drift and diffusivity evaluated with the discussed data-driven approach (points) and corresponding fits (solid lines). Left: $\beta \simeq +0.11$. Right: $\beta \simeq -0.10$. Drift terms are fitted with sinusoidal functions, diffusivities with constant values. Parameters of the simulations: $M=8$, $m=1$, $k=0.5$, $N=600$, $n=15$. }\label{fig:posneg}
\end{figure}
 Fig.~\ref{fig:posneg} shows the drift and diffusivity terms as reconstructed by our approach. We considered two cases, one at positive and one at negative $\beta$.
 The drift term $F(P)$ is clearly proportional to $\partial_P K(P)$, as it should. The diffusivity is almost constant. Relation~\eqref{eq:einstein_gen} holds both at positive and at negative temperature. In particular, as it is clear from the figure, passing from positive to negative $\beta$ the proportionality factor between $F(P)$ and $K(P)$ changes its sign.
\begin{figure}
  \centering
  \includegraphics[keepaspectratio=true, width=0.48\linewidth]{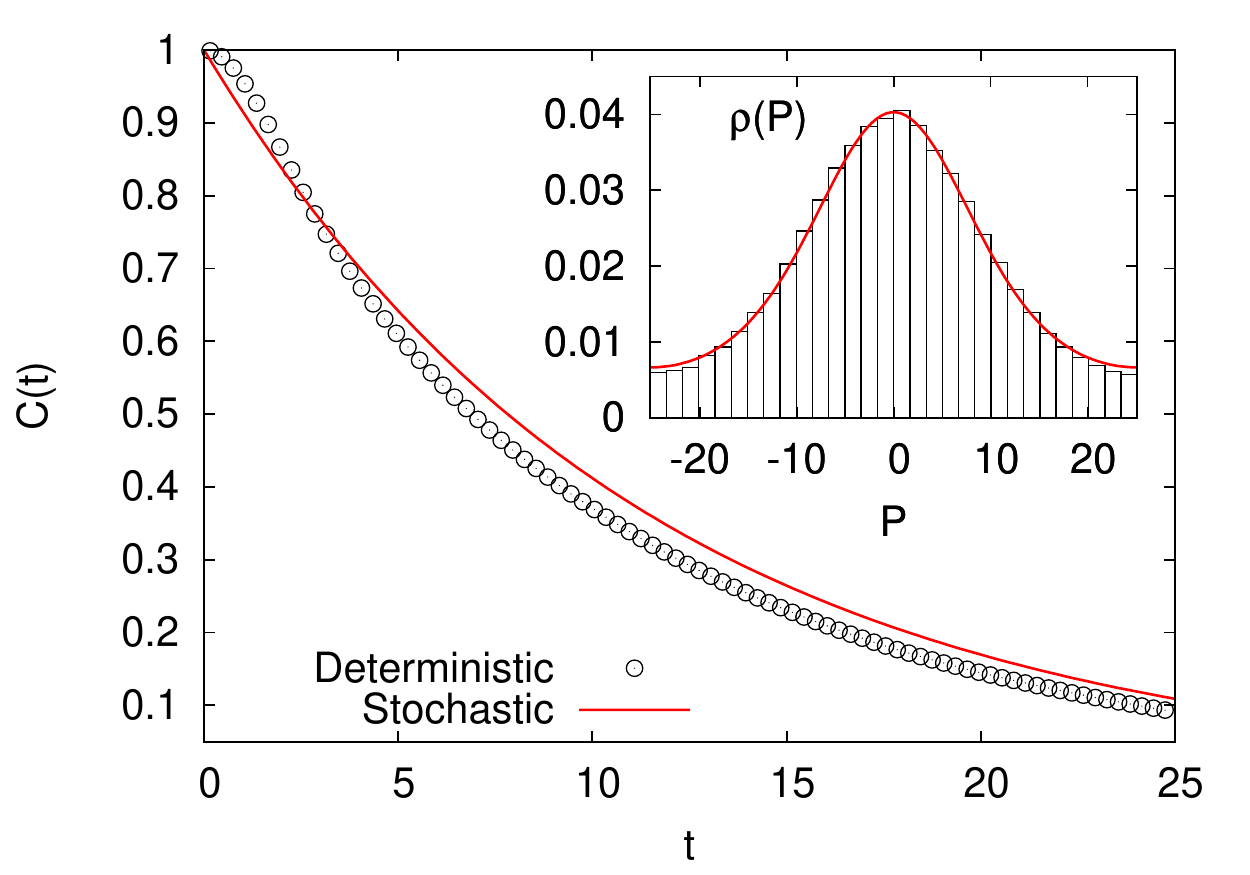}
   \includegraphics[keepaspectratio=true, width=0.48\linewidth]{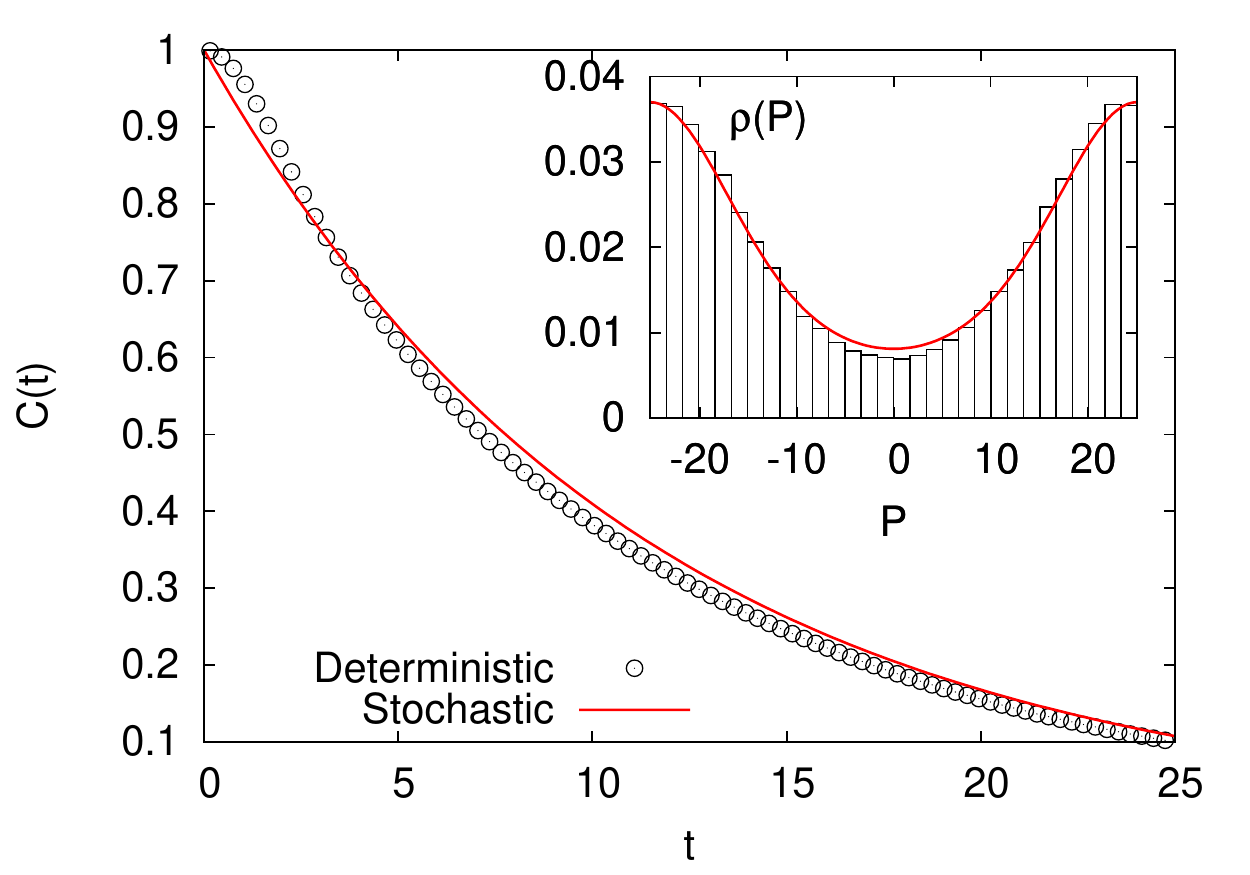} 
   \caption{Autocorrelation functions of the velocity in the original dynamics (circles) and in the coarse-grained model (solid lines) at $\beta=2$ (left) and $\beta=-2$ (right). In the insets, probability density functions of $P$ in the two cases, both for the original dynamics (boxes) and for the reconstructed model (solid lines).} \label{fig:autoc}
\end{figure}
\\
Finally, we can look at the steady probability density functions and at the autocorrelations of $P$, in order to compare the ones computed with the coarse-grained model and those from the original deterministic system. As shown in Fig.~\ref{fig:autoc}, the agreement is quite good, both at positive and negative $\beta$.
\\
The fact that it is possible to write down a Langevin-like equation with $\beta<0$, which properly reproduces the behavior of a slow degree of freedom subjected to the action of a bath, sounds quite relevant in the context of the long-lasting debate about negative temperatures. Several Authors propose the adoption of a definition of entropy alternative to Eq.~\eqref{eq:beta}, the so-called ``volume entropy'' or ``Gibbs entropy'' $S_G$~\cite{dunkel14}. A statistical description based on $S_G$ is able to reproduce the classical results of thermodynamics \textit{exactly}, even if the number of degrees of freedom of the system is small~\cite{romero13, dunkel14, hilbert14, campisi15}: due to this remarkable property, $S_G$ may appear as the right mechanical analogue of the thermodynamic entropy. However, since $S_G$ is by definition a non-decreasing function of energy, $\beta$ would never be negative. We stress that with this alternative choice it would not be possible, in our opinion, to give a coherent description of the above results.

\subsection{ Some remarks on the method}
\begin{figure}
  \centering
  \includegraphics[keepaspectratio=true, width=0.5\linewidth]{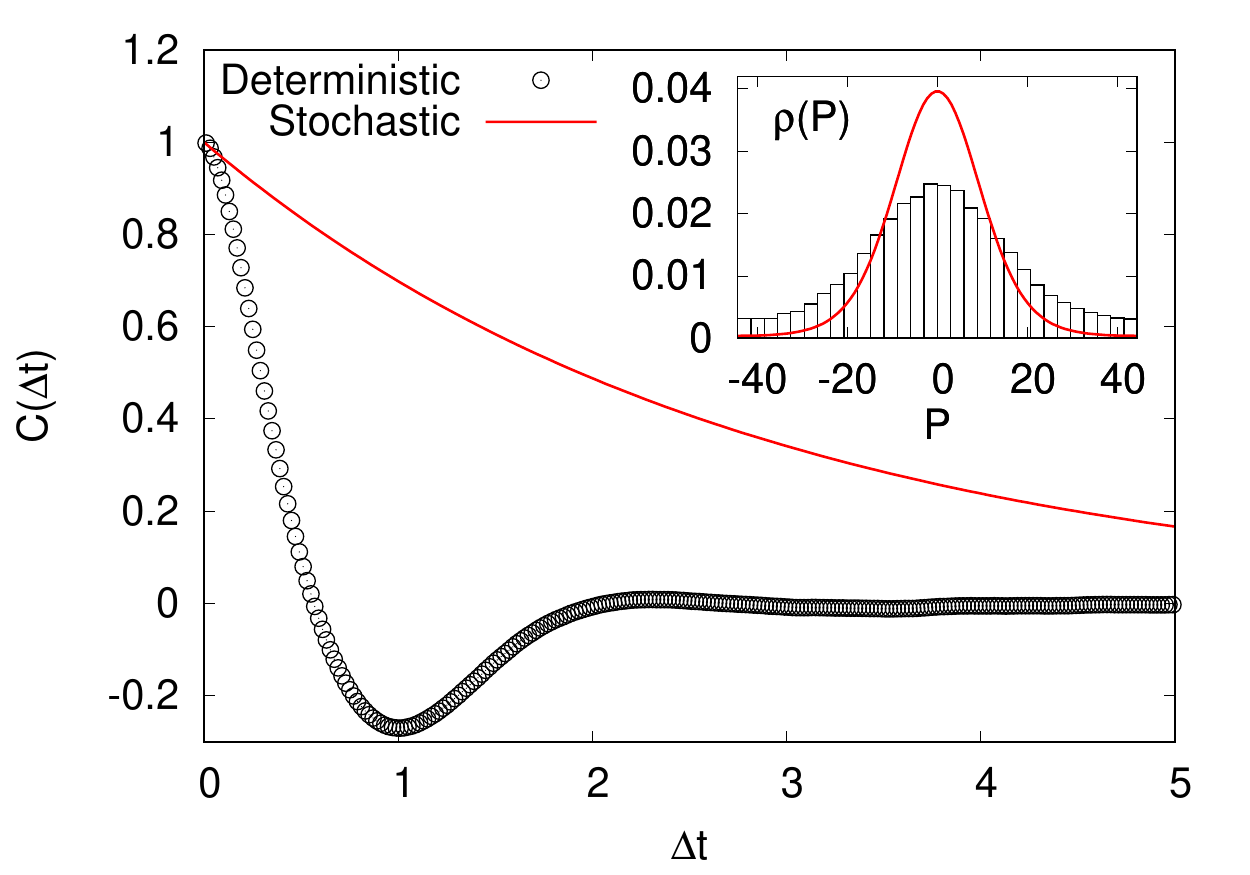}
   \caption{A case in which the described method fails: the autocorrelation function of the original system~\eqref{eq:fakeH} (circles) and that of the reconstructed model (solid red line) do not coincide. Similarly, in the inset, the original (boxes) and the reconstructed p.d.f. (solid red line) are different. Paramters: $M=200$, $m=1$, $k=2500$, $2N=2000$, $\beta=1.04$.} \label{fig:contro}
\end{figure}

At a first glance it seems that everything is quite easy.
On the other hand one has to face several difficulties:
\begin{enumerate}
 \item Time-scales separation  does not imply Markovianity.
 \item It is true that, in some cases, adding variables can 
lead to a Markovian process 
(e.g. if colored noise is present); unfortunately
 a general method to find the ``right'' variables does not exist.
Such a difficulty has been expressed in a rather vivid way
in the caveat of Onsager and Machlup~\cite{onsager53}:
   {\it How do you know you have taken enough variables, 
for it to be Markovian?}
\item The  procedure discussed in this Section cannot be applied ``blindly'',
and Markovianity should always be checked a posteriori.
\end{enumerate}
In order to give an example of the above troubles and
how mathematics can help to select/eliminate models 
we discuss the case shown in Fig.~\ref{fig:contro}, which is obtained from a system with Hamiltonian
\begin{equation}
\label{eq:fakeH}
 H=\sum_{i=-N}^{N} [1-\cos(p_i/\sqrt{m_i})] + k\sum_{i=-N}^{N+1}[1-\cos(q_{i}-q_{i-1})]
\end{equation}
where $(p_0,q_0)\equiv(P,Q)$ and $m_i=m+\delta_{i,0}(M-m)$.
\\
Also in this case one can expect that the presence of a large ``mass'' would lead to a scale-separation between the dynamics of the slow and those of the fast particles. However let us note that the additivity condition mentioned above is not satisfied in this case.
In such a situation, in spite of 
time-scales separation, it is impossible to
find a $1D$ Langevin equation for $P$.
The reason of such impossibility is  a property of the Fokker-Planck operator: in $1D$ equilibrium Langevin equation the autocorrelation function 
cannot be negative~\cite{gardiner85}. Therefore if we want to build an effective Langevin equation,  
it is necessary to introduce (at least) another variable.

\section{Data driven models: LE for a granular system}
\label{sec:granular}

\begin{figure}
  \centering
  \includegraphics[keepaspectratio=true, width=0.48\linewidth]{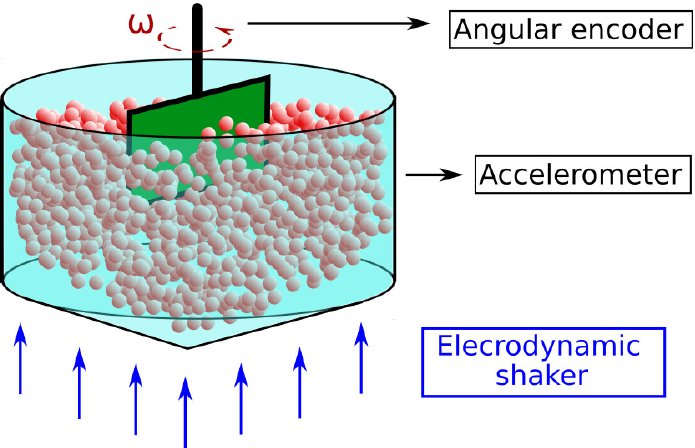}
      \caption{Sketch of the set-up for the experiment discussed in Ref.~\cite{scalliet15}. The system is composed of a large number of beads immersed in a cylindrical container, which is shaken from below with a fixed frequency. A blade is immersed in the granular gas and it is free to rotate along the vertical axis. Its angular position is measured by an encoder. The acceleration felt by the container is also measured.} \label{fig:scheme}
\end{figure}

Let us now discuss how to use the  procedure  for the building of effective
equations when just  experimental data are available,
in  a system  for which we have not  a well established 
theoretical understanding.
\\
The analyzed time series is that of the angular velocity of a rotator suspended
 in a vibrofluidized granular medium, as discussed in Ref.~\cite{scalliet15}. In Fig. \ref{fig:scheme} we show a sketch of the experiment set-up. We recall that in a granular medium kinetic energy is not conserved and therefore Hamiltonian modelling discussed before is not applicable.
\\
The granular medium is composed of $N$ spheres of diameter $d=4$ mm,
placed in a cylindrical container whose volume is $\sim$ 7300 times that of a single sphere. 
The container is vertically shaken with a signal whose spectrum is approximately flat in a range 
between  $f_{min}= 200$ Hz and $f_{max}= 400$ Hz. 
A blade, a "massive tracer", with cross section $\sim 8 d \times  4 d$, is suspended into the granular 
medium and rotates around a vertical axis. 
Its angular velocity $\omega(t)$ and the traveled angle of rotation $\theta(t)=\int_0^t \omega(t')dt'$
are measured with a time-resolution of $2$ kHz.

\subsection{The Gas limit}

\begin{figure}
  \centering
  \includegraphics[keepaspectratio=true, width=1.\linewidth]{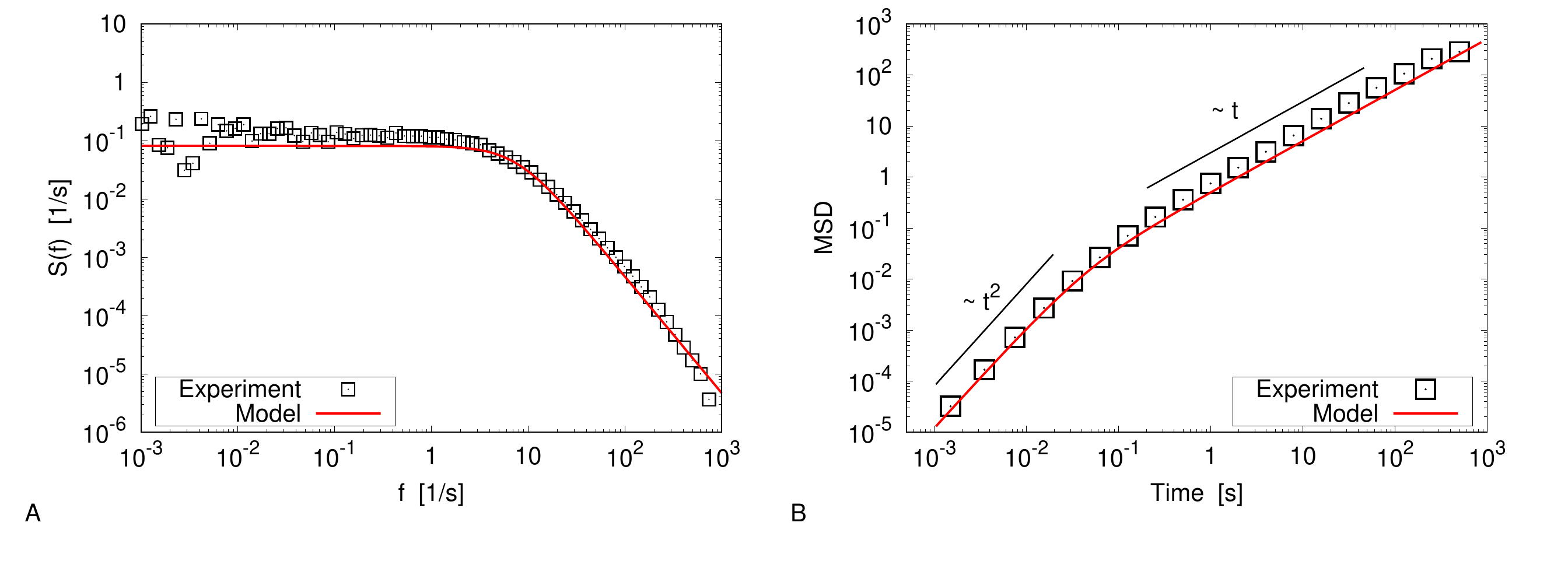}
     \caption{Velocity power density spectrum (left) and mean square displacement (right) for the angular position of the blade in the dilute-gas limit. Squares represent experimental data, solid lines are computed with the reconstructed model.} \label{fig:gran_dilute}
\end{figure}

In the dilute limit, e.g. with $N=350$, and
 packing fraction  $\sim 5 \%$, the simplest possible scenario holds.
The blade performs  a standard rotational Brownian motion, and a $1D$
linear Langevin equation for $\omega$ is enough for a good description of the observed behavior.
We have
$$
{d \omega \over dt}= -{\omega \over \tau} + c \eta\,,
$$
where the parameters $\tau$ and $c$ can be easily obtained with the procedure
discussed in Sec.~\ref{sec:testing}, see Fig.~\ref{fig:gran_dilute}. This analysis has been done in Ref.~\cite{baldogran}.

\subsection{The Cold Liquid limit}
\begin{figure}
  \centering
  \includegraphics[keepaspectratio=true, width=1.\linewidth]{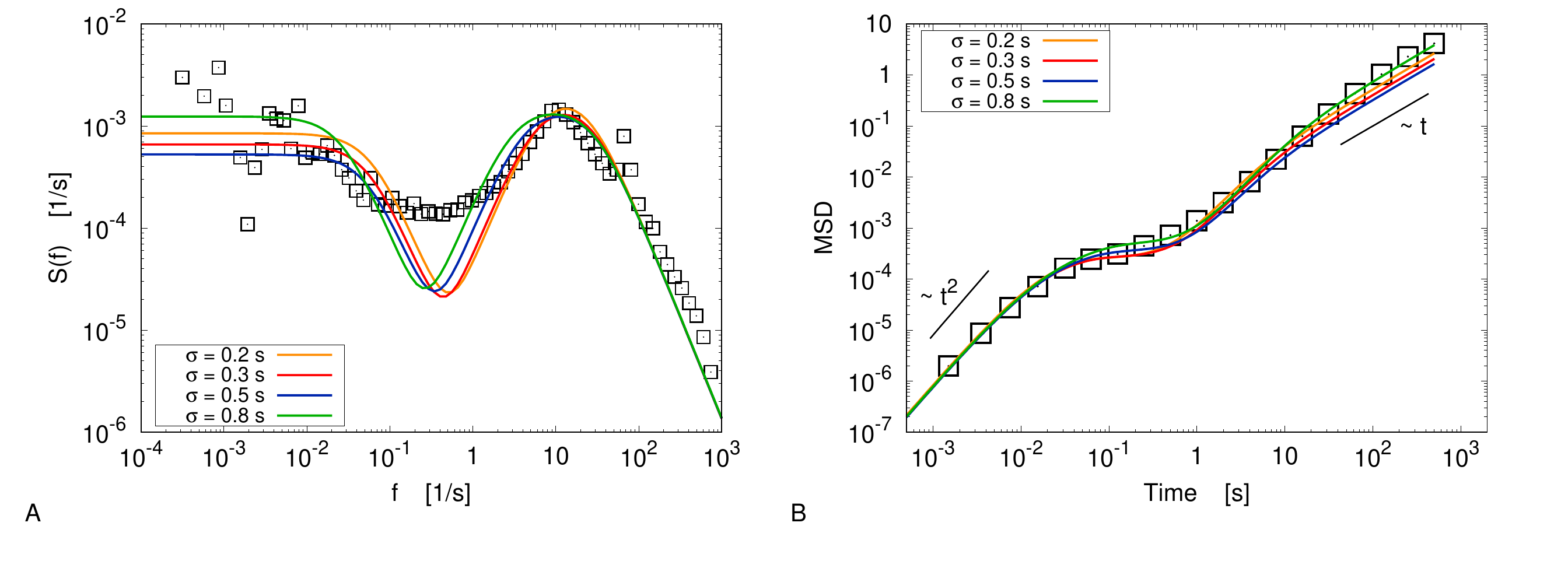} 
   \caption{Velocity power density spectrum (left) and mean square displacement (right) for the angular position of the blade in the cold-liquid limit. Squares represent experimental data, solid lines are computed with the reconstructed model. Different colors of the lines correspond to differnt values of the parameter $\sigma$ used for the analysis, as mentioned in the text.} \label{fig:gran_dense}
\end{figure}

On the contrary, in the dense regime, e.g.
with $N=2600$ and  packing fraction  $ \sim 30 \%$, the 
scenario of the usual standard rotational Brownian motion 
fails, and a rich phenomenology appears (e.g., a superdiffusive behavior at long time-scales).
A single  equation for $\omega$ is not able to describe in a proper way the observed results.
It is necessary to introduce, at least,  a second variable.
These new variable, which  describes the slow behavior of the probe,
has been obtained by performing a running average with a Gaussian window function
$$
\theta_0(t)={1 \over \sqrt{2 \pi \sigma^2}} \int e^{-{(t-t')^2 \over 2 \sigma^2}} \theta(t') \, dt'
$$
the range of values in which $\sigma$ should be chosen is suggested by the shape of $S(f)$.
\\
The reconstructed Langevin equation for the variables
$$
 \theta \,, \,  \theta_0 \,, \,  \omega={d\theta \over dt} \,, \,  \omega_0={d\theta_0 \over dt}\,,
$$
discussed in Ref.~\cite{baldogran}, is able to reproduce the main statistical features, including the anomalous
diffusion, see Fig~\ref{fig:gran_dense}. 
\\
\\
\\
\\
\section{ About enthusiasm for Big Data,  Artificial Intelligence
and Machine Learning }
\label{sec:bigdata}
Recently the possibility of extracting knowledge by data mining (i.e.,
through the algorithmic analysis of large amounts of data) seems to
suggest the emergence of a fourth paradigm, a new scientific
methodology.
In particular a consistent fraction of the scientific community seems close
 to conclude that it is possible to
build models with  black box protocols.
\\
On the other hand, in our opinion, the past experience show that 
for the problem of the prediction, there are rather severe
limitations for  such an approach.
For instance in Refs.~\cite{cecconi12,hosni18} the reader can find a detailed discussion
about the
good reasons (mainly due to  the Kac's lemma on the recurrence time) to be
skeptical about the data-centric enthusiasm supporting a general
philosophy starting from "raw data", without constructing modeling
hypotheses and, therefore, without theory.
\\
Let us discuss again how the ability of choosing the "right" 
variables typically requires a conceptual abstraction which is key 
to scientific discoveries. 
\\
In the '80s, some researchers in the field of artificial intelligence
(AI) devised BACON, a computer program to automatize scientific
discoveries.  Apparently, BACON was able to "discover" the third
Kepler's law~\cite{bradshaw71,grabiner86}.  
\\ Let us look at the details of the procedure used by
BACON, and by Kepler: 
\begin{itemize}
 \item BACON used as input the numerical values
of distance from the Sun, $D$, and revolution period, $P$, of
planets. The program, then, discovered that $D^3$ is proportional to
$P^2$.  
\item For Kepler the raw observables were not $D$ and $P$, but
a list of planetary positions seen from the Earth at different
times. In his discovery, Kepler chose the "right" variables $D$ and
$P$ as he was guided by strong beliefs in mathematical harmonies as
well as the (at that time) controversial theory of Copernicus.
\end{itemize}
Is it appropriate to claim that  AI methods can replace the traditional creative 
approach to scientific discoveries~\cite{buchanan19}?

\section{Summary and conclusions}
\label{sec:conclusion}

Let us briefly summarize the main points discussed in this paper,
and try to fix some conclusions.
\begin{enumerate}
 \item LE can be generalized to cases with ``unusual'' kinetic energy, and 
 in particular to systems which can show negative temperature.
\item In order to check its validity, one can simulate a  bath  and find the effective LE,
 this approach works in presence of:
 \begin{itemize}
  \item time-scale separation;
  \item Markovianity of the variable we are looking at.
 \end{itemize}
\item A procedure to find LE for the slow dynamics can be used in the treatment of data if
 one is able to identify the relevant variables.
\item The choice of the "good variables" does not follow from a mechanical protocol,
 it can be only suggested by intuition and/or a preliminary understanding of
 the main aspects of the phenomena under investigation.
\end{enumerate}
We conclude with a remark about the machine learning methods (MLM) and AI in the research.
It is matter of fact that in several cases MLM and AI had been able to succeed;
we believe that  community  have a great opportunity to give a contribute in the  
understanding of  the range of applicability of MLM and AI.

\begin{acknowledgments}
 {We aknowldge fruitful collaboration and useful discussions with F. Borra, F. Cecconi, M. Cencini, M. Falcioni, A. Gnoli, A. Prados, A. Puglisi and A. Sarracino.}
\end{acknowledgments}

\section*{Bibliography}

\bibliographystyle{unsrt}
\bibliography{biblio}

\end{document}